\documentclass[showpacs,superscriptaddress,preprintnumbers,amsmath,amssymb,11pt]{revtex4}
\usepackage{graphicx}
\usepackage{epsfig}  
\usepackage{dcolumn}
\usepackage{bm}
\usepackage{color}

\def\gev{{\rm GeV}}

\begin{document}

\title{Tension between scalar/pseudoscalar new physics contribution \\
to $B_s \rightarrow \mu^+ \mu^-$ and 
$B \rightarrow K \mu^+ \mu^-$}

\author{Ashutosh Kumar Alok}
\affiliation{Tata Institute of Fundamental Research, Homi Bhabha
Road, Mumbai 400005, India}
    
\author{Amol Dighe}
\affiliation{Tata Institute of Fundamental Research, Homi Bhabha
Road, Mumbai 400005, India}
                                                          
\author{S. Uma Sankar}
\affiliation{Indian Institute of Technology Bombay, Mumbai-400076, 
India}


\begin{abstract} 

New physics in the form of scalar/pseudoscalar operators cannot 
lower the semileptonic branching ratio $B(B \rightarrow K \mu^+ \mu^-)$
below its standard model value.
In addition, we show that the upper bound on the leptonic branching ratio
$B(B_s \rightarrow \mu^+ \mu^-)$ sets a strong constraint on the
maximum value of $B(B \rightarrow K \mu^+ \mu^-)$ in models with
multiple Higgs doublets:
with the current bound, $B(B \rightarrow K \mu^+ \mu^-)$ 
cannot exceed the standard model prediction by more than 2.5\%.
The conclusions hold true even if the new physics couplings are complex.
However these constraints can be used to restrict new 
physics couplings only if the theoretical and experimental errors 
in $B(B \rightarrow K \mu^+ \mu^-)$ are reduced to a few per cent.
The constraints become relaxed in a general class of models with
scalar/pesudoscalar operators.
\end{abstract} 

\pacs{13.20.He, 12.60.-i}


\maketitle

One of the major aims of the large hadron collider (LHC),
about to start operating soon, is to look for Higgs particles 
within and beyond the standard model (SM).
Even a direct observation of a Higgs particle
will not suffice to tell us whether it is the SM Higgs or not.
An understanding of possible scalar/pseudoscalar
new physics (SPNP) interactions through indirect means
is therefore extremely crucial.

The flavor changing neutral interaction $b \rightarrow s \mu^+ \mu^-$ 
serves as an important probe to test
higher order corrections to the SM as well as 
to constrain many new physics models. 
This four-fermion interaction is responsible for the purely 
leptonic decay $B_s \rightarrow \mu^+ \mu^-$, for the
semileptonic decays $B \rightarrow (K, K^*) \mu^+ \mu^-$
and also for the radiative leptonic decay 
$B_s \to \mu^+ \mu^- \gamma$.
The semileptonic decays have been experimentally observed at 
BaBar and Belle \cite{babar-03, belle-03, babar-06, hfag}.  
The pseudoscalar semileptonic decay has the branching ratio
\begin{equation}
B(B\rightarrow K \mu^{+} \mu^{-})  = 
(4.2^{+0.9}_{-0.8})\times10^{-7} \; , 
\label{br-kmumu} 
\end{equation}
which has been obtained with $\sim 350$ fb$^{-1}$ of data.
These values are consistent with the SM predictions
\cite{bobeth,ali-02,lunghi,kruger-01},
and the experimental errors are expected to reduce to 
$\sim 2\%$ at the forthcoming Super-B factories \cite{hitlin}. 
At the moment there is about 20\% uncertainty in these SM 
predictions due to the error in the quark mixing matrix element 
$V_{ts}$ and the uncertainties related to strong interactions.
Improvements in the lattice calculations and the
measurement of $V_{ts}$ are likely to bring this error
down to a few per cent within the next decade.

The purely leptonic decay $B_s \rightarrow \mu^+ \mu^-$ is highly
suppressed in the SM, 
the prediction for its branching ratio
being $(3.35\pm 0.32)\times10^{-9}$ \cite{buras-01}. 
The uncertainty in the SM prediction is mainly due to the
uncertainty in the decay constant $f_{B_s}$ and $V_{ts}$. 
This decay is yet to be observed in experiments.
Recently the upper bound on its branching ratio 
has been improved to \cite{cdf-07}
\begin{equation}
B(B_s \rightarrow \mu^+ \mu^-) < 0.58 \times10^{-7}
\quad  (95\% ~{\rm C.L.}) \; ,
\label{mumu-lim}
\end{equation}
which is still more than an order of magnitude 
away from its SM prediction.
The decay $B_{s}\rightarrow\mu^{+}\mu^{-}$ will
be one of the important rare B decay channels to be studied 
at the LHC and 
we expect that the sensitivity of about $10^{-9}$ can be reached 
in a few years \cite{Lenzi:2007nq}.

In the context of these decays, one needs to focus only on 
new physics from scalar/pseudoscalar interactions, since
(i) new physics in the form of vector/axial-vector operators is 
highly constrained by the data on 
$B \rightarrow (K, K^*) \mu^+ \mu^-$ as shown in 
\cite{alok-sankar01}, and 
(ii) new physics in the form of tensor and magnetic dipole 
operators does not contribute to $B(B_s \rightarrow \mu^+ \mu^-)$.
A measured value of $B(B_s \rightarrow \mu^+ \mu^-) \gtrsim 10^{-8}$
indicates that the new physics 
must be in the form of scalar/pseudoscalar operators. 

We take the effective Lagrangian for the four-fermion transition 
$b \rightarrow s \mu^+ \mu^-$ to be \cite{bobeth}
\begin{equation}
L(b \rightarrow s \mu^{+} \mu^{-})  =  L_{SM} + L_{SP} \; , 
\end{equation}
where
\begin{widetext}
\begin{equation}
L_{SM}  =  \frac{\alpha G_F}{\sqrt{2} \pi} V_{tb} V^\star_{ts} \biggl\{ C^{\rm eff}_9(\bar{s} \gamma_\mu P_L b)\,
\bar{\mu} \gamma_\mu \mu  + C_{10}(\bar{s} \gamma_\mu P_L b)\,\bar{\mu} \gamma_\mu \gamma_5 \mu +
2 \frac{C^{\rm eff}_7}{q^2} m_b \, (\bar{s} i \sigma_{\mu\nu} q^\nu P_R b) \, \bar{\mu} \gamma_\mu \mu \biggr\} \; ,
\label{SML}
\end{equation}
\begin{equation}
L_{SP}  =  \frac{\alpha G_F}{\sqrt{2} \pi} V_{tb} V^\star_{ts} \biggl\{
\tilde{R}_S \, (\bar{s}\,P_R\, b) \, \bar{\mu} \,\mu + 
\tilde{R}_P \, (\bar{s}\,P_R\, b) \, \bar{\mu} \gamma_5 \mu \biggr\} \; .
\label{LSP}
\end{equation}
\end{widetext}
Here $P_{L,R} = (1 \mp \gamma_5)/2$ and $q$ is the sum of the 
$\mu^+$ and $\mu^-$ momenta. 
$\tilde{R}_S$ and $\tilde{R}_P$ are the scalar and pseudoscalar 
new physics couplings respectively, which in general can be complex. 
We use the notation $\tilde{R}_S \equiv R_S e^{i\delta_S},
\tilde{R}_P \equiv R_P e^{i\delta_P}$.
Here the phases are restricted to be $0 \leq (\delta_S, \delta_P) < \pi$, 
whereas $R_S$ and $R_P$ can take positive as well as negative values. 
Within SM, the Wilson coefficients in Eq.~(\ref{SML}) 
have the following values \cite{bobeth}:
\begin{equation}
C_{7}^{\rm eff} = -0.310 \; , \; 
C_{9}^{\rm eff} = +4.138 + Y(q^2) \; , \; C_{10} = -4.221,
\end{equation}
where the function $Y(q^2)$ is given in \cite{buras-95}.
These coefficients have an uncertainty of about $5 \%$,
which arises mainly due to their scale dependence.

In Eq.~(\ref{LSP}), we have taken only $P_R$ in the quark bilinear,
while the most general Lagrangian must have a linear combination 
of $P_L$ and $P_R$.
Here we start by considering the simpler case because 
SPNP operators mostly arise due to multiple Higgs doublets. 
In such models, the coefficient of $P_R$ in the Lagrangian 
is much larger than that of $P_L$ \cite{bobeth}. 
In two Higgs doublet model, for instance, the coefficient of 
$P_L$ is smaller by a factor of $m_s/m_b$ \cite{nierste}. 
We shall examine the consequences of considering
the most general quark bilinear in the latter part of this Letter.

In the following, we consider the interrelations between
the contributions of $L_{SP}$ to the branching ratios of 
the decays $B_s \to \mu^+ \mu^-$ and $B \to K \mu^+ \mu^-$. 
The effect of SPNP couplings on additional observables related 
to these decays, viz. forward-backward asymmetry in the 
semileptonic decay and the polarization asymmetry in the 
leptonic decay, has been studied in ~\cite{afb-alp}.
The contribution of $L_{SP}$ to $B \to K^*
\mu^+ \mu^-$ is so small \cite{bobeth} that no worthwhile 
correlation can be established between it and other decays. 
Also, $L_{SP}$
does not contribute to the radiative leptonic decay $B_s \to \mu^+ \mu^-
\gamma$ \cite{src-gaur, alok-sankar02}.

We first consider the contribution of $L_{SP}$ to the 
decay rate of  $B_{s}\rightarrow\mu^{+}\mu^{-}$.
The branching ratio is given by
\begin{eqnarray}
B_{SP}(B_{s}\rightarrow \mu^{+}\mu^{-}) & = & 
\frac{G_{F}^{2}\alpha^{2} m_{B_s}^{3} \tau_{B_s}}{64\pi^{3}}
|V_{tb}V_{ts}^{*}|^{2} f_{B_{s}}^{2} (R_{S}^{2} + R_{P}^{2}) \; .
\end{eqnarray}
Taking $f_{B_s}=(0.259 \pm 0.027) ~\gev$ \cite{mackenzie}, we get
\begin{equation}
B_{SP}(B_{s}\rightarrow \mu^{+}\mu^{-})\,=\,(1.43 \pm 0.30) \times
10^{-7}\, (R_{S}^{2} + R_{P}^{2}) \; .
\label{br_lep}
\end{equation}

\begin{figure*}
\begin{center}
\epsfig{file=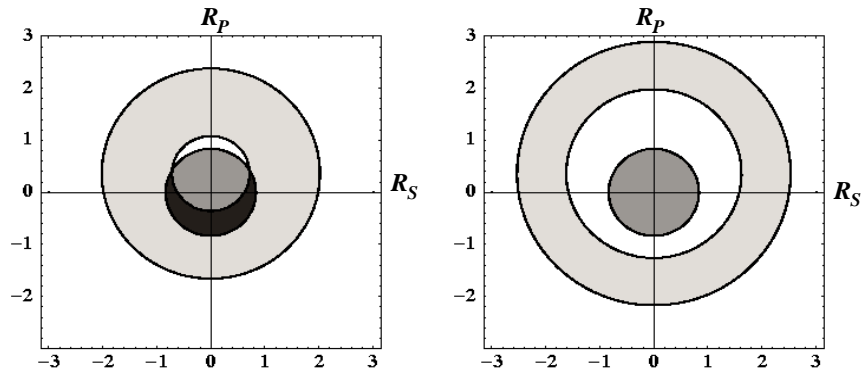,width=4.5in}
\caption{The allowed ranges of $R_S$ and $R_P$, when the new
physics couplings are real. In both figures, 
the dark grey circles centered at origin represent the  regions
allowed by the current $2\sigma$ upper bound on $B(B_s \to \mu^+ \mu^-)$.
The light grey annulus in each figure represents the parameter space
allowed by $B(B \to K \mu^+ \mu^-)$ at $2\sigma$.
The width of the annulus corresponds to the sum of the theoretical
and experimental errors, both of which are taken to be 2\%.
In the left panel, we take  $B(B \to K \mu^+ \mu^-) = (5.64 \pm 0.11) 
\cdot 10^{-7}$. 
The overlap between the allowed regions is represented by the black crescent.
In the right panel we take $B(B \to K \mu^+ \mu^-) = (6.04 \pm 0.12 ) \cdot 10^{-7}$,
where the allowed parameter spaces do not overlap.
}
\label{agree}
\end{center}
\end{figure*}

Note that the present experimental upper limit on 
$B(B_s \rightarrow \mu^+ \mu^-)$ is an order of magnitude 
larger than the SM prediction. 
In the following, 
we will assume that the SPNP will provide an order of magnitude
increase of $B(B_s \rightarrow \mu^+ \mu^-)$. 
In such a situation, the SM amplitude 
can be neglected in the calculation of the branching ratio. 
Equating the expression in Eq.~(\ref{br_lep}) to the present
95\% C.L. upper limit in Eq.~(\ref{mumu-lim}), 
we get the inequality
\begin{equation}
(R_{S}^{2} + R_{P}^{2}) \leq 0.70 \; ,
\label{lepconst}
\end{equation}
where we have taken the $2\sigma$ lower bound for the 
coefficient in Eq.~(\ref{br_lep}).
Thus, the allowed region in the $R_S$--$R_P$ parameter
space is the interior of a ``leptonic'' circle of radius 
$r_\ell \approx 0.84$
centered at the origin, as indicated in 
both the panels of Fig.~\ref{agree}.
As the upper bound on $B(B_s \rightarrow \mu^+ \mu^-)$
goes down, the radius of the circle will shrink.

We now turn to the semileptonic decay 
$B \rightarrow K \mu^+ \mu^-$.
The measured branching ratio is consistent with the SM prediction,
though there is a $25\%$ error in the measurement and 
about $20\%$ error in the theoretical prediction
due to uncertainties in  $V_{ts}$, form factors and 
Wilson coefficients (which in turn depend on $V_{ts}$).
With the addition of the SPNP contribution,
the theoretical prediction for the
net branching ratio becomes \cite{bobeth}
\begin{eqnarray}
B(B \rightarrow K \mu^+ \mu^-) & = &  
\left[5.25 + 0.18(R_S^2 + R_P^2) 
-0.13 R_P \cos\delta_P \right]  (1 \pm 0.20) \times 10^{-7} \; ,
\label{brsl}
\end{eqnarray}
In Eq.~(\ref{brsl}), the first term is purely due to the SM, the
second term is purely due to SPNP and the third term is due to the 
interference of the two. 
The theoretical errors arise from one tensor and two vector 
form factors in the SM, and a scalar form factor in SPNP
(which is related to one of the SM vector form factors).
We have made the simplifying
assumption that the fractional uncertainties in all the form 
factors are the same. 

Eq.~(\ref{brsl}) can be rewritten as
\begin{equation}
B(B \rightarrow K \mu^+ \mu^-) =   (1 + \epsilon) B_{\rm SM}  \; ,
\label{epsilon}
\end{equation}
where $\epsilon$ is the fractional change in the branching ratio
due to SPNP.
The maximum negative value that $\epsilon$ can take is $-0.005$,
thus implying that the SPNP new physics cannot lower the branching 
ratio $B(B \rightarrow K \mu^+ \mu^-)$ by more than 0.5\% 
below its standard model value.
Indeed, if the theoretical and experimental errors in this quantity 
were improved to 5\%, with the central values unchanged, 
the discrepancy cannot be accounted for by SPNP at $2\sigma$.

Let us first consider the case where the new couplings $R_S$ and $R_P$
are real, which is typical for the class of models where the only
charge-parity violation comes from the CKM matrix elements.
Using Eqs.~(\ref{br-kmumu}) and (\ref{brsl}), we get
\begin{equation}
R_S^2 + (R_P-0.36)^2  =
\frac{B_{\rm exp}}{(0.18 \pm 0.036) \times
10^{-7}} - 29.04 \; ,
\label{kmumu-constraint}
\end{equation}
where $B_{\rm exp}$ is the measured value of $B(B_s \to K \mu^+ \mu^-)$.
The region in the $R_S$--$R_P$ plane allowed by the 
measurement of $B(B_s \to K \mu^+ \mu^-)$ is then an 
``semileptonic'' annulus
centered at $(0,0.36)$, as shown in both the panels
of Fig.~\ref{agree}.
The inner and outer boundaries of this region correspond to
the lower and upper bounds of the right hand side of
Eq.~(\ref{kmumu-constraint}).
The right hand side turns out to be negative
if $B_{\rm exp}$ is below the SM prediction
by more than 0.5\%.
Then the radius of the circle becomes imaginary, which implies
that the discrepancy of the measurement with the SM
cannot be explained by SPNP.

To illustrate the tension between the quantities 
$B(B_s \to \mu^+ \mu^-)$ and $B(B \to K \mu^+ \mu^-)$,
we consider the scenario where the errors in both 
$B_{\rm SM}$ and $B_{\rm exp}$ have been reduced to 2\%,
while keeping the upper limit on $B(B_s \to \mu^+ \mu^-)$
at its current value.
The allowed $R_S$--$R_P$ parameter space is shown
in Fig.~\ref{agree}.
If the lower limit on $B_{\rm exp}$
is small enough, the semileptonic annulus will overlap with
leptonic circle, as shown in the left panel.
However, if the lower limit on $B_{\rm exp}$
is larger than a critical value
(determined by the bound on the leptonic branching ratio),
then there is no region of overlap as shown in the right panel.
In such a situation, the difference between 
$B_{\rm exp}$ and $B_{\rm SM}$ cannot be accounted for by SPNP 
because of the constraint coming from the leptonic mode.

We represent the radius of the leptonic circle by 
$r_\ell$ and the inner (outer) radius of the semileptonic
annulus by $r_{in}$ ($r_{out}$).
There is tension between the two measurements if
\begin{equation}
r_{in} - r_\ell > 0.36 \; , 
\label{condition}
\end{equation}
in which case the regions allowed by the two branching
ratios do not overlap.
Given the current value of $r_\ell = 0.84$, 
we require $0< r_{in} < 1.2$ for an overlap.
This implies that the $2\sigma$ lower limit on
$B_{\rm exp}$ should be between $4.93 \times 10^{-7}$
and $5.67 \times 10^{-7}$.
(We have added the theoretical and experimental errors
in quadrature.)
If the upper bound on $B(B_s \to \mu^+ \mu^-)$ is improved
by a factor of 5, the $2\sigma$ range for the lower limit
on $B_{\rm exp}$ would be $(4.93-5.57) \times 10^{-7}$.
For the tension to be manifest in future experiments, 
the reduction of errors in $B_{\rm exp}$ and $B_{\rm SM}$ 
is the most crucial.

When $\tilde{R}_S$ and $\tilde{R}_P$ are complex, the constraint 
Eq.~(\ref{kmumu-constraint}) becomes
\begin{equation}
R_S^2 + (R_P - 0.36 \cos \delta_P )^2  =
\frac{B_{\rm exp}}{(0.18 \pm 0.036) \times
10^{-7}} - 29.17 + (0.36 \cos\delta_P)^2 \; .
\label{kmumu-constraint-complex}
\end{equation}
For nonzero $\delta_P$, the center of the semileptonic annulus
shifts along the $R_P$ axis, while the radius of the annuli
are almost unchanged.
If the allowed regions do not overlap for $\delta_P =0$
(as illustrated in the right panel of Fig.~\ref{agree}),
then they will not overlap for any value of $\delta_P$.
Hence the tension between 
$B(B_s \to \mu^+ \mu^-)$ and $B(B \to K \mu^+ \mu^-)$ persists,
and gives rise to the same constraints on the semileptonic
branching ratio even if the SPNP couplings are complex.

In writing the effective SPNP Lagrangian in 
Eq.~(\ref{LSP}), we considered only the quark bilinear 
$\bar{s} P_R b$. Lorentz Invariance of the Lagrangian also
allows the bilinear $\bar{s} P_L b$ in general. We can take
this generalization into account by replacing $\bar{s} P_R b$
by $\bar{s} ( \alpha P_L + P_R) b$, where $\alpha$ is the 
strength of the $\bar{s} P_L b$ bilinear relative to that of
$\bar{s} P_R b$. 
With this modification, $B(B \to K \mu^+ \mu^-)$ 
is driven by the sum of the two quark bilinears with different 
chiralities, whereas $B(B_s \to \mu^+ \mu^-)$ 
depends on their difference \cite{hiller}.
The expressions for the branching ratios of the two 
processes considered here are:
\begin{eqnarray}
B(B_s \rightarrow \mu^+ \mu^-) & = & 
(1-\alpha)^2(R_{S}^{2} + R_{P}^{2}) \,
(1.43 \pm 0.30)  \times 10^{-7}  ,
\label{alpha-leptonic} \\
B(B \rightarrow K \mu^+ \mu^-) & = &
\left[5.25 + 0.18 \, (1+\alpha)^2\,(R_S^2 + R_P^2) 
-0.13 \, (1+\alpha) R_P  \right] \, (1 \pm 0.20 )
\times 10^{-7}  .
\label{alpha-semileptonic}
\end{eqnarray}
Here we have taken $R_S$,$R_P$ and $\alpha$ to be real for simplicity.
For $\alpha=0$, Eqs.~(\ref{alpha-leptonic}) and 
(\ref{alpha-semileptonic}) reduce to Eqs.~(\ref{br_lep}) and
(\ref{brsl}) respectively.
For the special case $\alpha = 1$, the new physics has 
no contribution to $B_s \rightarrow \mu^+ \mu^-$ because 
the quark bilinear is pure scalar and the corresponding
pseudoscalar meson to vacuum transition matrix element is zero.
In such cases, $B(B_s \rightarrow \mu^+ \mu^-)$ is entirely
due to the SM, and provides no constraints on 
$B(B \rightarrow K \mu^+ \mu^-)$.

\begin{figure}
\begin{center}
\epsfig{file=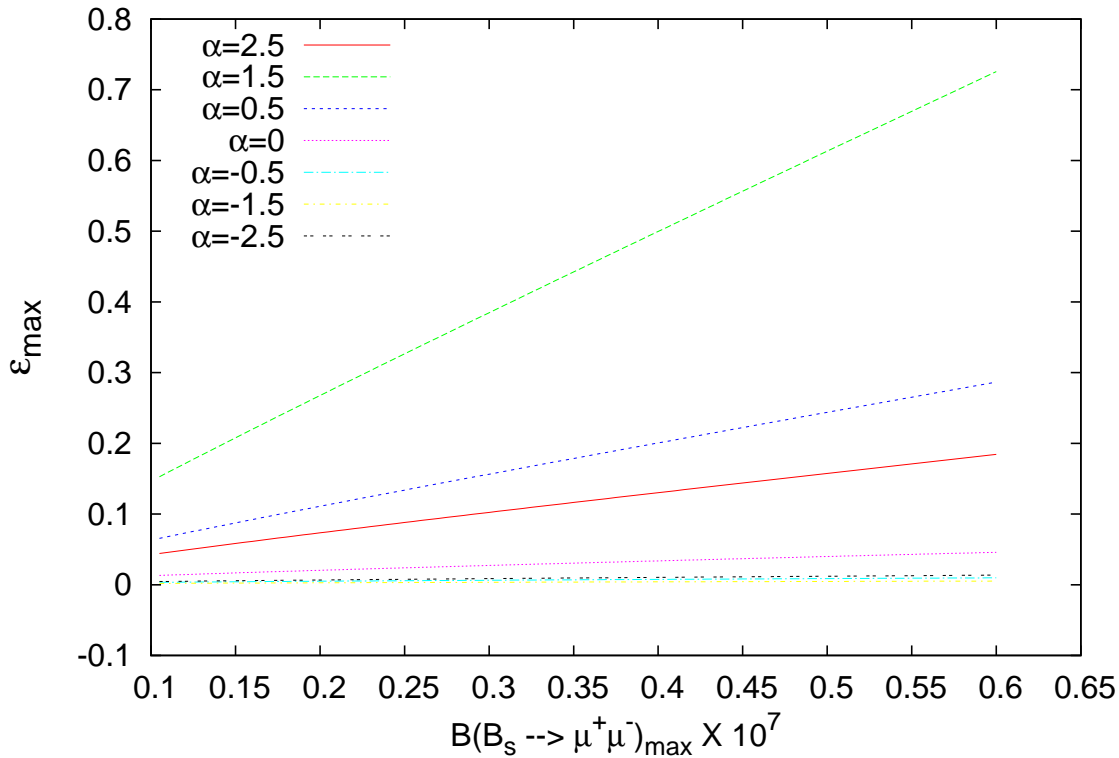,width=6in}
\caption{$\epsilon_{\rm max}$ as a function of
the $2\sigma$ upper bound on $B(B_s \to \mu^+ \mu^-)$ 
for different values of $\alpha$.
We have taken the theoretical error on $B(B_s \to \mu^+ \mu^-)$ 
to be 20\%; decreasing it would further constrain $\epsilon_{\rm max}$. 
\label{alpha-plot}}
\end{center} 
\end{figure}

In Fig.~\ref{alpha-plot}, we show $\epsilon_{\rm max}$,
the maximum fractional deviation of $B(B \rightarrow K \mu^+ \mu^-)$ 
from its SM prediction as defined in Eq.~(\ref{epsilon}), as a function
of the $2\sigma$ upper bound on $B(B_s \rightarrow \mu^+ \mu^-)$.
The minimum allowed value of $\epsilon$ is almost independent
of the value of $\alpha$ and the leptonic upper bound, 
and is approximately $-0.005$.
For the class of models with multiple Higgs doublets, 
$\alpha=0$, and the maximum value of $\epsilon$
is restricted to $+0.025$, as seen in earlier discussions.
With the additional freedom generated by the extra parameter
$\alpha$, this severe constraint is relaxed.
For example, for the models with $\alpha \approx 1.5$,
the value of $\epsilon$ may be as large as $+0.7$, as can be
seen in the figure.
In general for positive $\alpha$ values, $\epsilon_{\rm max}$ increases 
with $\alpha$ for $\alpha <1.0$, and decreases thereafter.
When $\alpha <0$,  Eq.~(\ref{alpha-leptonic}) indicates that
the constraints on $R_S$ and $R_P$ should become more restrictive.
As a result, $\epsilon$ is constrained to be even smaller.
From the figure, $\epsilon_{\rm max}$ for
negative $\alpha$ are seen to be very close to zero, and 
the corresponding $\epsilon_{\rm max}$ curves are almost 
overlapping. This implies that for negative $\alpha$, any 
significant deviation of $B(B \rightarrow K \mu^+ \mu^-)$ from SM 
is impossible with SPNP.

For the measurements of $B(B_s \rightarrow \mu^+ \mu^-)$ and
$B(B \rightarrow K \mu^+ \mu^-)$ to be compatible with SPNP,
the lower bound on $B(B \rightarrow K \mu^+ \mu^-)$ 
should be less than $(1+\epsilon_{\rm max}) B_{\rm SM}$. 
Thus, the upper bound on $B(B_s \rightarrow \mu^+ \mu^-)$ 
and the lower bound on $B(B \rightarrow K \mu^+ \mu^-)$ allow us to
constrain the value of $\alpha$ in a class of models
that involve new physics scalar/pseudoscalar couplings.

In this letter, we have parameterized scalar/pseudoscalar
new physics in terms of the effective operators given 
in Eq.~(\ref{LSP}). In general, the introduction of new 
scalar/pseudoscalar fields into a model leads to not only new
effective operators but also modification of the coefficients
of the SM operators, e.g. the Wilson coefficients 
 $C_7$, $C_9$ and $C_{10}$ shown in Eq.~(\ref{SML}). 
However, it has been shown that these modifications due to new 
scalar/pseudoscalar fields are very small \cite{nierste,dai-huang-huang}.
We have computed these changes in the two Higgs doublet model 
and found them to be at most $1\%$. Thus, our assumption of
retaining the SM values for the Wilson coefficients, even in the presence
of new scalar/pseudoscalar fields, is valid.

In summary, we have shown that in a class of models with 
new scalar/pseudoscalar operators, which includes models with
multiple Higgs doublets,
the SPNP couplings are strongly constrained by the upper bound on 
$B(B_s \rightarrow \mu^+ \mu^-)$, and in turn restrict
the allowed values of $B(B \rightarrow K \mu^+ \mu^-)$
to within a narrow range around its SM prediction.
Future precise measurements of these two branching ratios
have the potential not only to give an evidence for new physics, but
also to reveal the nature of its Lorentz structure.
However in order to achieve this, the theoretical as well as
experimental errors on $B(B \rightarrow K \mu^+ \mu^-)$
need to be reduced to a few per cent.

\acknowledgments

We thank the organizers of WHEPP-X at IMSc, Chennai
where preliminary discussions on this topic started.
We also thank G. Hiller for useful comments.


\end{document}